\documentclass[twocolumn,showpacs,
amsmath,amssymb]{revtex4}
\usepackage{amssymb}
\usepackage[dvips]{graphicx}
\begin{document}

\title{ Low-temperature  proximity effect in clean metals}
\author{A. S. Alexandrov$^{1,2}$ and  V. V. Kabanov$^{2,1}$}

\affiliation{$^1$Department of Physics, Loughborough University,
Loughborough LE11 3TU, United Kingdom\\
$^2$Josef Stefan Institute 1001, Ljubljana, Slovenia }

\begin{abstract}
Theories of  proximity effect in layered superconductor-normal metal
(SN) structures usually deal with  a hypothetic normal metal with no
interaction between electrons and with finite temperatures often
close to the superconductor critical temperature. We present an
asymptotic solution of the Gor'kov equations in the opposite
low-temperature limit for a clean normal metal with a repulsive
interaction between electrons. The order parameter in the metal
exhibits a power-law decay, $\Delta(x)\propto \xi/x$, as a
function of the distance from the SN boundary, $x$,  with a
proximity length $\xi$ strongly depending on the repulsive
interaction.

\end{abstract}


\maketitle

In recent years  investigations of different SN structures
\cite{degennes} have gone through a vigorous revival.  In
particular, superconductor/ferromagnet structures \cite{buzdin,tag},
cuprate SNS junctions \cite{bozp,ale}, and mesoscopic SN structures
\cite{lambert,schon,zaikin,mota} have been experimentally studied
and addressed theoretically.

The superconducting order parameter penetrates into a bulk normal
metal across the SN boundary. The microscopic theory of this
proximity effect has been developed  at finite temperatures (for
reviews see \cite{degennes2,likharev}), using  the Eilenberger
formalism \cite{eilen} and the semiclassical Usadel approximation
\cite{usa} for solving  the Gor'kov equations \cite{gor}  close to and
below $T_c$, in particular in the dirty limit \cite{likharev2}. The
clean case has been studied by Falk \cite{falk} for a hypothetic
normal metal with no interaction between electrons, and by Maksimov
and Potapenko \cite{max} for a
 "normal" metal with a weak \emph{attractive} interaction and
reduced critical temperature. An exponential decay of $\Delta(x)
\propto \exp(-x/\xi)$, as a function of the distance from the SN
boundary, has been found with the proximity length $\xi \propto 1/T$
in the clean limit \cite{max}, and $\xi \propto 1/\sqrt{T}$ in the
dirty limit \cite{degennes,degennes2,likharev}. To the best of our
knowledge a role of the \emph{repulsive} interaction in the normal
metal has not been addressed in the clean case, while its effect on
the proximity length and the Josephson current has been found
marginal in the dirty case \cite{likharev2}.

Here we present an asymptotic solution of the Gor'kov equations for
the SN boundary between a bulk  superconductor and   a bulk \emph{clean}
normal metal with the \emph{repulsion} between electrons  at low
temperatures, Fig.1.

It is convenient to fourier-transform the Matsubara normal, ${\cal G}_{\omega}({\bf
r-r'},x,x')$, and anomalous, ${\cal F}^{+}_{\omega}({\bf
r-r'},x,x')$, Green's functions (GFs) along the boundary, ${\bf r }=\{y,z\}$
with the wave vector  ${\bf k}=\{k_y,k_z\}$,

 \begin{equation}
 {\cal G}_{\omega}({\bf
r-r'},x,x')=(2\pi)^{-2}\int d{\bf k} {\cal
G}_{\omega,k}(x,x')\exp[i{\bf k \cdot (r-r')}] \nonumber
\end{equation}

 \begin{equation}
 {\cal F}^{+}_{\omega}({\bf
r-r'},x,x')=(2\pi)^{-2}\int d{\bf k} {\cal
F}^{+}_{\omega,k}(x,x')\exp[i{\bf k \cdot (r-r')}]. \nonumber
\end{equation}
 The Gor'kov equations are derived using equations of motion for the Matsubara operators as
\begin{equation}
{1\over{2m}}\left(a^2+{{\partial}^2\over{{\partial}x^2}}\right){\cal
G}_{\omega,k}(x,x') +\Delta(x) {\cal
F}^{+}_{\omega,k}(x,x')=\delta(x-x'), \label{G}
\end{equation}
\begin{equation}
{1\over{2m}}\left(a^{*
2}+{{\partial}^2\over{{\partial}x^2}}\right){\cal
F}^{+}_{\omega,k}(x,x') -\Delta^{*}(x) {\cal G}_{\omega,k}(x,x')=0,
\label{F}
\end{equation}
\begin{equation}
{1\over{2m}}\left(a^{2}+{{\partial}^2\over{{\partial}
x'^2}}\right){\cal G}_{\omega,k}(x,x') +\Delta^{*}(x') {\cal
F}^{+*}_{-\omega,k}(x,x')=\delta(x-x'), \label{G'}
\end{equation}
\begin{equation}
{1\over{2m}}\left(a^{*2}+{{\partial}^2\over{{\partial}x'^2}}\right){\cal
F}^{+*}_{-\omega,k}(x,x') -\Delta(x') {\cal
G}_{\omega,k}(x,x')=0. \label{F'}
\end{equation}
Here $m, k_F$ are  the electron effective  mass and the
Fermi-momentum, respectively, which are taken the same in the
superconductor and in the normal metal for mathematical
transparency, $a^2=2m (i\omega -\xi_k)$ with $\xi_k=(k^2-k_F^2)/2m$
and $ a=sign(\omega)\sqrt{m(\rho+\xi_k)}+i \sqrt{m(\rho-\xi_k)}$, $\rho\equiv
+\sqrt{\omega^2+\xi_k^2}$, $\omega = 2\pi T(n+1/2)$ is the Matsubara
frequency ($n=0, \pm 1, \pm 2,...$). We use $\hbar=k_B=1$ here
and below, and ${\cal
F}^{+*}_{-\omega,k}(x,x')={\cal
F}_{\omega,k}(x,x')$.
\begin{figure}
\begin{center}
\includegraphics[angle=-90,width=0.43\textwidth]{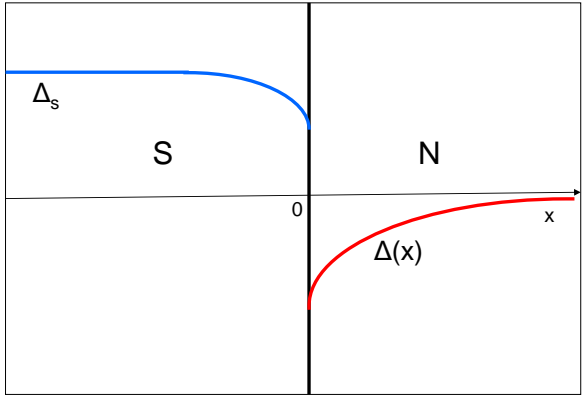}
\vskip -0.5mm \caption{ The order-parameter profile near the boundary between
a superconductor (S) and a normal metal (N) with the repulsive electron-electron interaction.}
\end{center}
\end{figure}
The superconducting order parameter
 \begin{equation}
\Delta(x)\equiv - V(x) T
(2\pi)^{-2} \sum_{\omega}\int d{\bf k} {\cal F}_{\omega,k}(x,x), \label{def}
\end{equation}
is  a solution of an integral equation,
\begin{equation}
\Delta (x)=-V(x)\int_{-\infty}^{\infty} dx' {\cal
G}^{(0)}_{-\omega,k}(x,x')\Delta(x'){\cal G}_{\omega,k}(x,x'),
\label{D}
\end{equation}
where $V(x)$ is the electron-electron contact interaction, which is
negative (attractive) in the superconductor at $x<0$, and positive
(repulsive) in the normal metal, $V(x)\equiv V_c >0$, at $x>0$,
Fig.1. The  GF of a bulk normal metal is
\begin{equation}
{\cal G}^{(0)}_{\omega,k}(x,x')={m \over{ia}} \exp(ia|x-x'|).
\end{equation}
The Gor'kov equations are supplemented by the boundary conditions,
$\Delta (-\infty) \equiv \Delta_s$, $\Delta(\infty)=0$, and all GFs
should be continuous with respect to $x,x'$ together with their
first derivatives at the boundary $x=0$ or $x'=0$.

To solve the integro-differential system of equations
\ref{G}-\ref{D} let us assume that the repulsive interaction in the
normal metal significantly reduces the order parameter, so that the
latter is small,  $\Delta(x) \ll \Delta_s$, far away from the
boundary at $x \gg \xi$, where the proximity length $\xi$ is small
compared with the superconductor coherence length, $\xi_s$, $\xi \ll
\xi_s$. That allows us to use in Eq.(\ref{D}) a solution, ${\cal
G}_{\omega,k}(x',x) \approx {\cal G}^{st}_{\omega,k}(x',x)$ of  the
Gor'kov equations \ref{G}-\ref{F'} with a step-like order parameter
$\Delta(x)=\Delta_s \Theta(-x)$, where $\Theta (x)$ is the Heaviside
step  function. This solution can be readily obtained  by matching
GFs and their derivatives at the boundaries between  4 domains: a superconductor domain $S$, where both arguments are negative $x,x'<0$, two mixed domains $M$ with $x<0<x'$ or $x'<0<x$, and a normal domain $N$, where $x,x'
>0$,  Fig.2.

In the normal domain the solution   is found as (see also \cite{max,falk})
\begin{equation}
{\cal
G}^{stN}_{\omega,k}(x,x')={\cal G}^{(0)}_{\omega,k}(x,x')+Ae^{ia(x+x')},
\label{step}
\end{equation}
\begin{equation}
{\cal F}^{+ stN}_{\omega,k}(x,x')=B e^{i(ax'-a^{*}x)},
\label{step2}
\end{equation}
and in one of the mixed domains, $x'<0<x$, as
\begin{eqnarray}
{\cal
G}^{stM}_{\omega,k}(x,x')&=&C{2m\Delta_s \over(b^2-a^2)}e^{i(ax-bx')}\cr
&+&D{2m\Delta_s\over(b^{*2}-a^2)}e^{i(ax+b^{*}x')},
\end{eqnarray}
\begin{equation}
{\cal F}^{+ stM}_{\omega,k}(x,x')=C^{*}e^{i(b^{*}x'-a^{*}x)}+D^{*}e^{-i(bx'+a^{*}x)},
\end{equation}
where $A,B,C$ and $D$ are constants and  $b^2=2m (i\epsilon-\xi_k)$, $\Im b >0 $,
$\epsilon=\sqrt{\omega^2+\Delta_s^2}$. The constants are found from ${\cal G}^{stN}_{\omega,k}(x,0)={\cal G}^{stM}_{\omega,k}(x,0)$, ${\cal F}^{+ stN}_{\omega,k}(x,0)={\cal F}^{+ stM}_{\omega,k}(x,0)$ and from ${\partial\over{\partial
x'}}{\cal G}^{stN}_{\omega,k}(x,x')={\partial\over{\partial
x'}}{\cal G}^{stM}_{\omega,k}(x,x')$, ${\partial\over{\partial
x'}}{\cal F}^{+ stN}_{\omega,k}(x,x')={\partial\over{\partial
x'}}{\cal F}^{+ stM}_{\omega,k}(x,x')$ at $x'=0$.

\begin{figure}
\begin{center}
\includegraphics[angle=-90,width=0.43\textwidth]{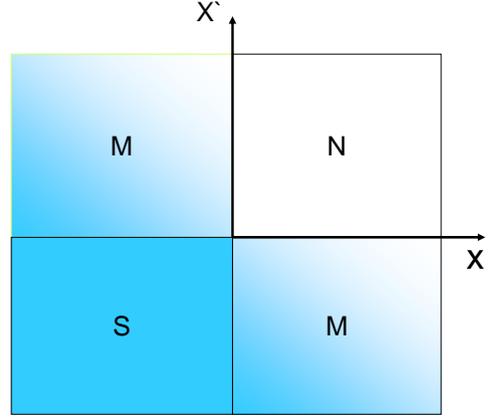}
\vskip -0.5mm \caption{ Four domains with different solutions for GFs with the step-like order parameter $\Delta(x)=\Delta_s \Theta(-x)$.}
\end{center}
\end{figure}

In particular we find
\begin{equation}
A={m\over{
ia}}\left( {2a ((a^{*}+b^{*}) (\epsilon +\omega)
+(a^{*}-b)(\epsilon-\omega))\over{|a+b|^2(\epsilon+\omega)+|a-b^{*}|^2(\epsilon-\omega)}}-1\right),
\end{equation} and
\begin{equation}
B=2m\Delta_s
{b+b^{*}\over{|a+b|^2(\epsilon+\omega)+|a-b^{*}|^2(\epsilon-\omega)}}.
\end{equation}

Integrating over the normal region, $x'>0$ in Eq.(\ref{D})    one can
keep only the first "normal" term of Eq.(\ref{step}) at sufficiently
large $x>\xi$, while integrating
 over the superconductor region, $x'<0$, one can
 use directly
 Eq.(\ref{step2}) and the definition of $\Delta(x)$, Eq.(\ref{def})  to obtain
\begin{eqnarray}
&&\tilde{\Delta}(x)=-V_c T m\sum_{\omega} \int {d{\bf
k}\over{(2\pi)^2}}\cr && \left[ u e^{-2x\Im a}+{m\over{|a|^2}}
\int_{0}^{\infty} dx' \tilde{\Delta}(x')e^{-2 |x-x'|\Im a}\right],
\label{integral}
\end{eqnarray}
where $\tilde{\Delta}=\Delta(x)/\Delta_s$ is the reduced order
parameter in the normal metal, and $u=B/m\Delta_s$.

\begin{figure}
\begin{center}
\includegraphics[angle=-0,width=0.50\textwidth]{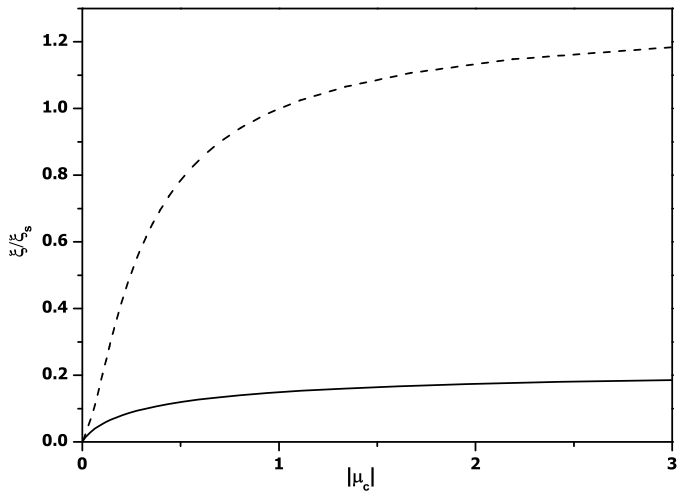}
\vskip -0.5mm \caption{SN proximity length $\xi$ (in units of the
superconductor coherence length) as a function of the repulsive pseudopotential
$\mu_c$ (solid line) and of the attractive potential (dashed line). }
\end{center}
\end{figure}
At finite temperatures and $x\gg v_F/(2\pi T)$ the main contribution
to the Matsubara sum in Eq.(\ref{integral}) comes from the $n=0$
term, so that the order parameter has the conventional exponential
asymptotic, $\Delta(x) \propto \exp(-2\pi T x/v_F )$, where
$v_F=k_F/m$ is the Fermi velocity. At sufficiently low and zero
temperatures one has $x \ll v_F/ (2\pi T)$ for any size of the
normal region, so that the exponential asymptotic is replaced by
some power decay \cite{likh}. To find the power we replace the
Matsubara sum by an integral over $\omega$ and integrating over
momentum and frequency in Eq.(\ref{integral}) obtain an
integral equation for the reduced order parameter, $\tilde{\Delta}(x)=\Delta(x)/\Delta_s$,
\begin{equation}
{\xi_s\over{x}}+\int_0^{\infty} dx'
{\tilde{\Delta}(x')\over{|x'-x|}}=-{\tilde{\Delta}(x)\over{\mu_c}},
\label{integral2}
\end{equation}
where $\mu_c=V_c mk_F/4\pi^2$ is the repulsion pseudopotential, and
$\xi_s=v_F/2\Delta_s$. One can  satisfy Eq.(\ref{integral2}) with a solution
decaying as inverse distance from the boundary,
\begin{equation}
\tilde{\Delta}(x)=-\xi/x. \label{solution}
\end{equation}
 Substituting Eq.(\ref{solution}) into Eq.(\ref{integral2}) yields
\begin{equation}
\tilde{\xi}={\mu_c\over{1+\mu_c\int_0^\infty dt/t|t-1|}}, \label{log}
\end{equation}
 where $\tilde{\xi}=\xi/\xi_s$ is the dimensionless proximity length.

The logarithmic divergency of the integral
in Eq.(\ref{log}) is an artifact of the step-function approximation, Eqs.(\ref{step},
\ref{step2}). We cut the divergency by excluding  small regions,  $t<t_{min} \ll 1$ and $|t-1|<t_{min}$ from the integral, where the step-function approximation  fails because $\tilde{\Delta}(x)$ becomes relatively large.  We chose $t_{min} =\tilde{\xi}$ since $\tilde{\xi}$ is the only  dimensionless parameter  in the normal region which corresponds to the cutoff of the integral over $x'$ in
Eq.(\ref{integral2}) at a cutoff length $x'>l_{min}=\tilde{\xi}x$, proportional to $x$. The result is a transcendental equation for
$\tilde{\xi}$,
\begin{equation}
\tilde{\xi}={\mu_c\over{1-3\mu_c\ln(\tilde{\xi})}}. \label{log2}
\end{equation}
The proximity length is shown in Fig.3 as a function of the
repulsion  $\mu_c$. It   strongly depends on the
repulsion with a maximum value $\tilde{\xi}_{max} \approx 0.22$ at
$\mu_c=\infty$ \cite{ref}. The magnitude of $\tilde{\xi}$ is
 small at any $\mu_c$, which
justifies our step-function approximation, Eqs.(\ref{step},
\ref{step2}) for solving the problem.

To verify the self-consistency of the approximation one can
estimate the correction, $\delta{\cal G}_{\omega,k}(x,x')$, to GF, Eq.(\ref{step})  due to the finite order-parameter
in the normal region, at large $x,x'>0$,
\begin{equation}
\delta{\cal G}_{\omega,k}(x,x')=-\int_{0}^{\infty} dx' {\cal
G}^{(0)}_{\omega,k}(x,x'')\Delta(x'') {\cal F}^{+ stN}_{\omega,k}(x'',x').
\end{equation}
Using Eq.(\ref{step2}) and Eq.(\ref{solution}) one obtains $\delta{\cal G}_{\omega,k}(x,x') \propto \tilde{\xi}\ln(x/l_{min})$, which is small as $\tilde{\xi}\ln(1/\tilde{\xi}) \ll 1$ at any $x$ with our choice of the cutoff length, $l_{min}=\tilde{\xi}x$. Another possible choice of the cutoff length $l_{min}=\xi$ does not  change the order parameter in a wide region $\xi_s<x< \xi_s/\tilde{\xi}$ because  the singularity is logarithmically weak. We note that
 the order parameter becomes
so small at very large distance from the boundary,  that the BCS mean-field approximation used in Eqs.
(\ref{G}- \ref{F'}) may break down \cite{ref2}.

While the order parameter saturates at a small value for any fixed $x$ in the normal region, the pair wave function, ${\cal
F}^N_{\omega,k}(x,x')$,  decreases as $1/\mu_c$ at large repulsion.  At first glance the existence of finite pair correlations in the repulsive normal metal looks  surprising because it  leads to a finite increase of the repulsive potential energy of the whole SN system. In particular, an individual  real-space pair of electrons bound by some attractive potential on one side of the boundary would stay at an infinite distance from the boundary with the repulsive interaction (zero proximity effect). However,  the electron-density  homogeneity  in  metals  creates a quantum pressure on pairs pushing them across the boundary. The reason for the failure of the pair wave function to die off  is that there is simply no pair-breaking mechanism for disrupting any correlation that drift across the boundary as in the case of a hypothetical normal metal with no interaction \cite{falk}. With the power law order parameter, $\Delta(x) \propto 1/x$, an increase of the potential energy, proportional to $\Delta^2(x)$ is finite, when it is integrated over the whole normal region. Compared with the step-like order parameter this increase is compensated by  a lowering of the kinetic energy near the  boundary.

It is instrumental to compare the SN proximity effect in the repulsive normal metal with the effect in a "normal" metal with a small attractive potential between electrons at low but finite temperatures above the transition temperature, $T_{cn}$ ($\ll T_c$), of the "normal" metal. Solving Eq.(\ref{integral2}) with a negative $\mu_c$ yields a positive order parameter
\begin{equation}
\tilde{\Delta}_{att}(x)=\xi_{att}/x \label{solution2}
\end{equation}
with the proximity length,
\begin{equation}
{\xi_{att}\over{\xi_s}}\equiv \tilde{\xi}_{att}={|\mu_c|\over{1+3|\mu_c|\ln(\tilde{\xi}_{att})}}, \label{log2}
\end{equation}
shown in Fig.3 as the dashed line. With increasing attraction the proximity length increases linearly as in the case of the repulsion, but much faster saturating at $\tilde{\xi}_{att}\approx 1.2$ for the values of the interaction which are well beyond the step-function (and BCS) approximations.  The power law decay of the order parameter,  Eq.(\ref{solution2}), holds for the region $ x  \ll v_F/(2\pi T)$, but different from the repulsive case only for finite temperatures, which are much higher than  $T_{cn}$ and smaller than $T_c$.

It is also worthwhile to mention that the normal Fermi-liquid with the hard-core repulsion between fermions becomes a p-wave \cite{kohn} or a d-wave \cite{gol} superfluid at very low temperatures (on the mK scale) due to the Friedel oscillations \cite{fri}  of the particle-particle interaction potential caused by screening. This unconventional pairing should not affect our results for the proximity effect with the conventional  s-wave superconductor as long as there is no symmetry-breaking mechanism. On the other hand if such a mechanism  is involved (e.g. the spin-orbit coupling),  the inhomogeneous s-wave order-parameter  can generate a secondary order-parameter of another symmetry \cite{kab,gor2,edel}. Finally,  the order-parameter in  a dirty normal metal should also exhibit some power-law decay at low temperatures, but the power could be different from $1/x$ of the clean metal \cite{kup}.

In summary we have solved a long-standing problem of the
low-temperature proximity effect in a clean normal metal with
repulsive interaction between electrons. We have found the
power-law decay  of the order parameter with the characteristic
proximity length  strongly reduced with respect to the
superconductor coherence length by the repulsive interaction in
the normal metal.

We thank  A. F. Andreev, A. A. Golubov, L. P. Gor'kov, D.
Khmelnitskii, M. Yu. Kuprianov ,  K. K. Likharev, L. A. Maksimov,
and A. V. Paraskevov for illuminating discussions and constructive
suggestions. The work was supported by EPSRC (UK) (grant no.
EP/D035589/1) and by the Slovenian Research Agency (ARRS) (grant
no. 430-66/2007-17).


\begin{thebibliography}{200}
\bibitem{ref} In fact, the value of  $\mu_c$ is restricted from above,
$\mu_c <1$, because of the Stoner ferromagnetic instability of the
normal metal at $\mu_c>1$.
\end{thebibliography}
\end{document}